\documentstyle[epsfig,psfrag,color]{article}

\begin{document}

\title{Topics in coarsening phenomena\\
{\small in Fundamental Problems in Statistical Physics XII}
{\small Leuven, Aug 30 - Sept 12, 2009}}
\author{Leticia F. Cugliandolo\\
Universit\'e Pierre et Marie Curie - Paris VI\\
Laboratoire de Physique Th\'eorique et Hautes Energies
}
\date{\today}

\maketitle

\begin{abstract}
These lecture notes give a very short introduction to 
coarsening phenomena and summarize some recent results in the field.
They focus on three aspects: the super-universality hypothesis, the 
geometry of growing structures, and coarsening in the spiral 
kinetically constrained model. 
\end{abstract}


\textcolor{black}
{\section{Introduction}}

These notes are complementary to my lectures at FPSP XII.  My aim is
to discuss in this text some recent developments in the theory of coarsening
phenomena~\cite{review-coarsening} (although I only covered some of
them in the talks).  Coarsening is possibly the simplest example of
macroscopic non-equilibrium relaxation and it has very far-reaching
practical applications.  This problem, although pretty old and rather
well understood qualitatively, is still far from having a close and
satisfactory quantitative description going beyond the pretty
successful but somehow deceiving use of the dynamic scaling
hypothesis.

After presenting the models I shall focus on and the coarsening
phenomenon, I shall discuss the following aspects: (1)
super-universality; (2) the statistics and geometry of evolving
ordered structures; (3) domain growth in a kinetically constrained
spin model, the two dimensional spiral model.  Some reasons why we
studied these problems are the following.  We wished to check the
super-universality hypothesis in more detail than what it had been
done so far. In our opinion, establishing the
validity of this hypothesis might be useful to eventually develop a
successful analytic method to compute scaling functions in coarsening
problems. We analyzed the geometry of domain growth with similar ideas
to the ones used in the study of critical equilibrium structures in
the past. Knowing in detail the time-dependent distribution of
structures and their geometric properties might give us hints into
what to search for in more complex systems with out of equilibrium
dynamics such as glasses.  In this respect, kinetically constrained
models are supposed to be toy models for glassy relaxation. We
demonstrated that the mechanism of relaxation in the spiral model is
of the coarsening type. This is interesting {\it per se} and for
applications to glassy physics too.  In short, apart from making our
understanding of coarsening more complete, we expect that these
results might be useful to better grasp the more complex dynamics of
glassy systems.

References to the original articles will be given in the main text. My
results on this field have been obtained in collaboration with the
colleagues and students that I warmly thank in the acknowledgements.

\textcolor{black}
{\section{Models}}

In this section we introduce three kinds of problems with phase
transitions: purely geometric models with percolation as the standard
example; models with an energy (or cost) function that have both
static and dynamic transitions, with Ising and Potts spin models as
typical instances; and models with purely dynamic phase transition
such as kinetically constrained spin systems focusing on the
bi-dimensional spiral model. Related to the latter case, I shall also
mention coupled map lattices as well as spin models with dynamic rules
that do not satisfy detailed balance.

\textcolor{black}
{
\subsection{Geometric and probabilistic models}}

We recall the definition of two purely geometric and probabilistic
models: site and directed percolation. These models change
behaviour at a special value of their control parameter, $p_c$, and
this threshold has many points in common with usual thermal
second-order phase transitions.  Percolation theory deals mainly with
the critical phenomenon and the study of cluster sizes and their
geometric properties.

\textcolor{black} {
\subsubsection{Site percolation}}

Percolation is the propagation of activities through connected space.
The site percolation model~\cite{percolation} is defined as
follows. For each site on the considered lattice one tosses a coin
independently. With probability $p$ the site is occupied and with
probability $1-p$ it remains empty.  A configuration is
constructed by sweeping the lattice once and filling the sites (or
not) independently with this rule.

Clusters are defined as groups of first-neighbour filled (or empty)
sites.  At a (lattice dependent) threshold $p_c$ an infinite cluster
(on an infinite lattice) appears. Strictly speaking, for $p \leq p_c$
($1-p \leq p_c$) there is no infinite connected-component of filled
(empty) sites with probability one while for $p>p_c$ ($1-p>p_c$) there
is an infinite connected-component of filled (empty) sites with
probability one. The critical values of $p_c$ in some typical
bi-dimensional lattices are: $p_c=1/2$ (triangular), $p_c\approx 0.59$
(square), $p_c\approx 0.62$ (honeycomb).

Percolation can be obtained as a limit of the Potts model 
(see its definition below) with $q\to 1$~\cite{Wu82}.

\textcolor{black}
{
\subsubsection{Directed percolation}}

Directed percolation~\cite{directed-percolation} is percolation with a
special direction along which the activity propagates.  It mimics
filtering of fluids through porous materials along a given
direction. For example, in the percolation of coffee making, the
source is the top surface of the grounded coffee that receives the
water and the activity is to have water flow. On a square lattice the
model is defined by assigning a directed bond 
on horizontal (say, pointing
to the left) and vertical (say, pointing down) edges with 
probability $p$.  This mimics different microscopic pore 
connectivity. This model displays a phase transition from a macroscopically
permeable (percolating) to an impermeable (non-percolating) state.  
Directed percolation on a $2d$ square lattice has a
transition at $p_c\simeq 0.705$.

More generally, the terms percolation and directed percolation stand for
universality classes of continuous phase transitions which are
characterized by the same type of collective behavior on large
scales. A number of critical exponents, linked to the size of the 
percolation cluster, correlation functions, and so on can be defined 
similarly to what is done in thermal critical phenomena~\cite{percolation}.

\textcolor{black}
{
\subsection{Hamiltonian systems}}

Physical models, as well as some mathematical problems such as those requiring
optimisation, have an energy or cost function associated to each
configuration. In equilibrium, the configurations are sampled with a
probability distribution function that depends on their energy and
the external parameters such as temperature or a chemical
potential. The experimental conditions, that is to say whether the
system is isolated or in contact with thermal or particle reservoirs,
dictate the statistical ensemble to be used (micro-canonical, canonical
or gran-canonical). Out of equilibrium, say when the system is let
evolve from an initial condition that is not one selected from the
equilibrium measure, the system wanders in phase space in a manner
that we shall discuss below.

\textcolor{black}
{
\subsubsection{The Ising model}}

{\it Anisotropic magnets} are modeled in a very simple way 
by using Ising spins, $s_i=\pm $, to describe the magnetic moments
and by choosing an adequate interaction among them. 
The Hamiltonian of the $d$ dimensional Ising model is 
\begin{equation}
H=-\sum_{\langle ij\rangle} J_{ij} s_i s_j 
\; . 
\label{eq:Ising}
\end{equation}
The spin variables are placed on the vertexes of a lattice and the sum
runs over nearest-neighbours.  The exchange interactions $J_{ij}$ are
all positive, favouring ferromagnetism but they can, in principle,
take different values.  We shall focus on the usual uniform case
$J_{ij}=J$, that is to say the \textcolor{black}{clean} model, and the
case in which the $J_{ij}$'s are \textcolor{black}{quenched random
variables} drawn from a probability distribution with positive
support, a \textcolor{black}{dirty} case. The latter defines the random-bond
Ising model (RBIM). One can also add a random field term to the
uniform model and thus construct the random field Ising model (RFIM).

A totally random spin configuration in which $s_i=\pm 1$ with 
probability a half is a realization of a site percolation  
configuration with $p=1/2$ and, in $d=2$, neither positive nor negative
clusters percolate. The identification is achieved by defining site
occupation variables $n_i=(s_i+1)/2=0,1$. 

\begin{figure}[t]
\centerline{
\input{FIGS/free-energy-vs-order-parameter-lowT.pslatex}
\hspace{0.25cm}
\input{FIGS/free-energy-vs-order-parameter-Tc.pslatex}
\hspace{0.25cm}
\input{FIGS/free-energy-vs-order-parameter-highT.pslatex}
\hspace{0.25cm}}
\vspace{-0.75cm}
\centerline{
\includegraphics[width=4cm]{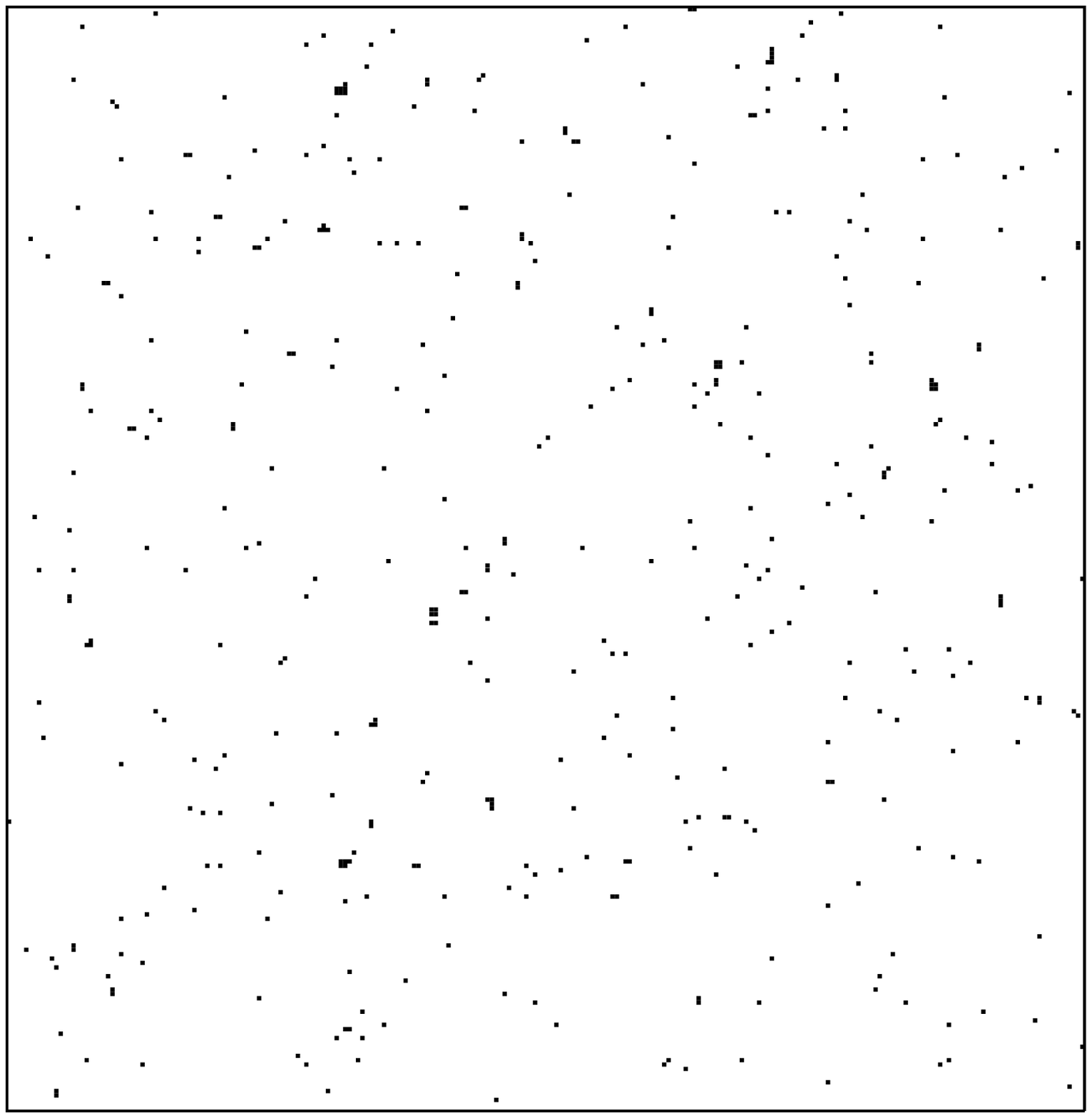}
\includegraphics[width=4.1cm]{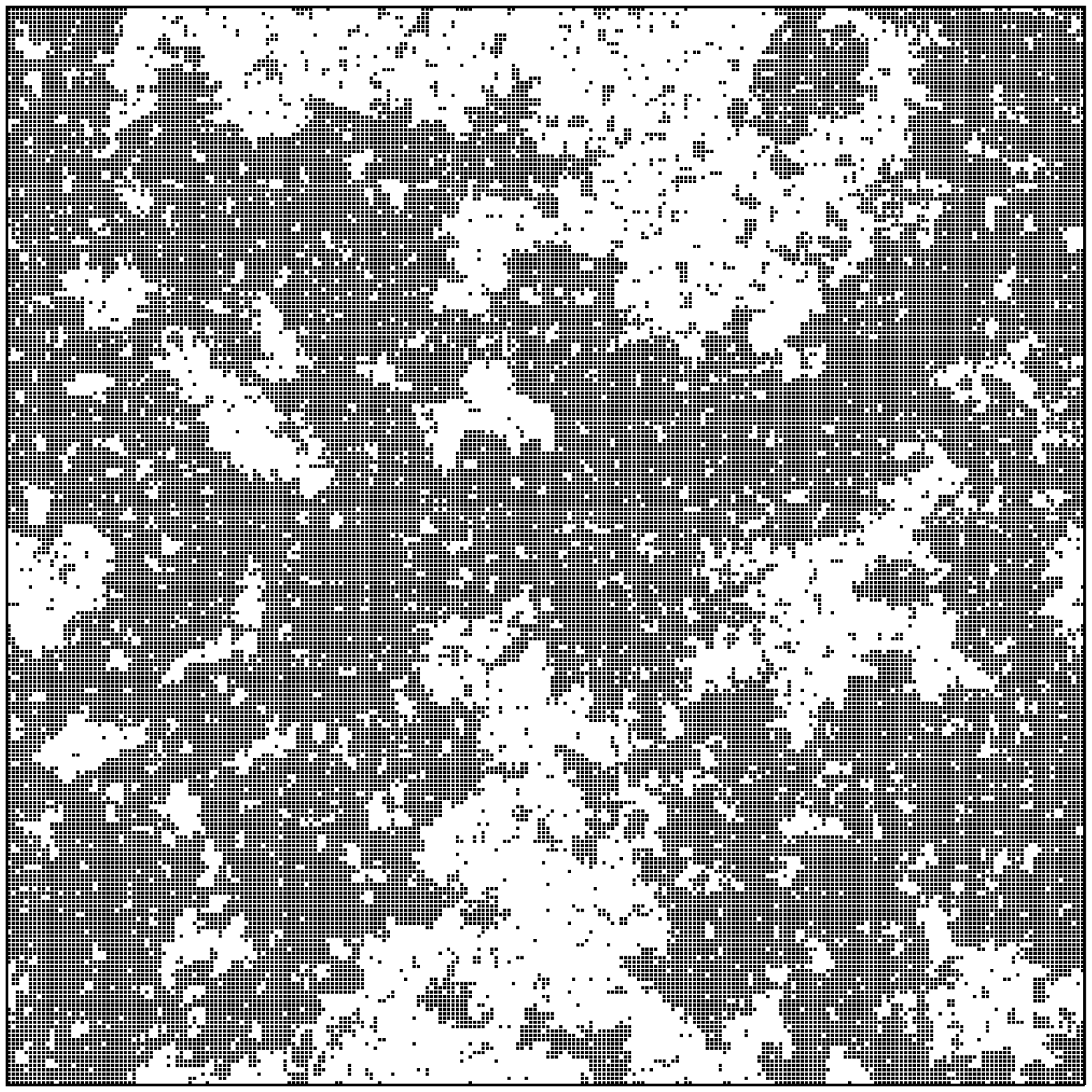}
\includegraphics[width=4.2cm]{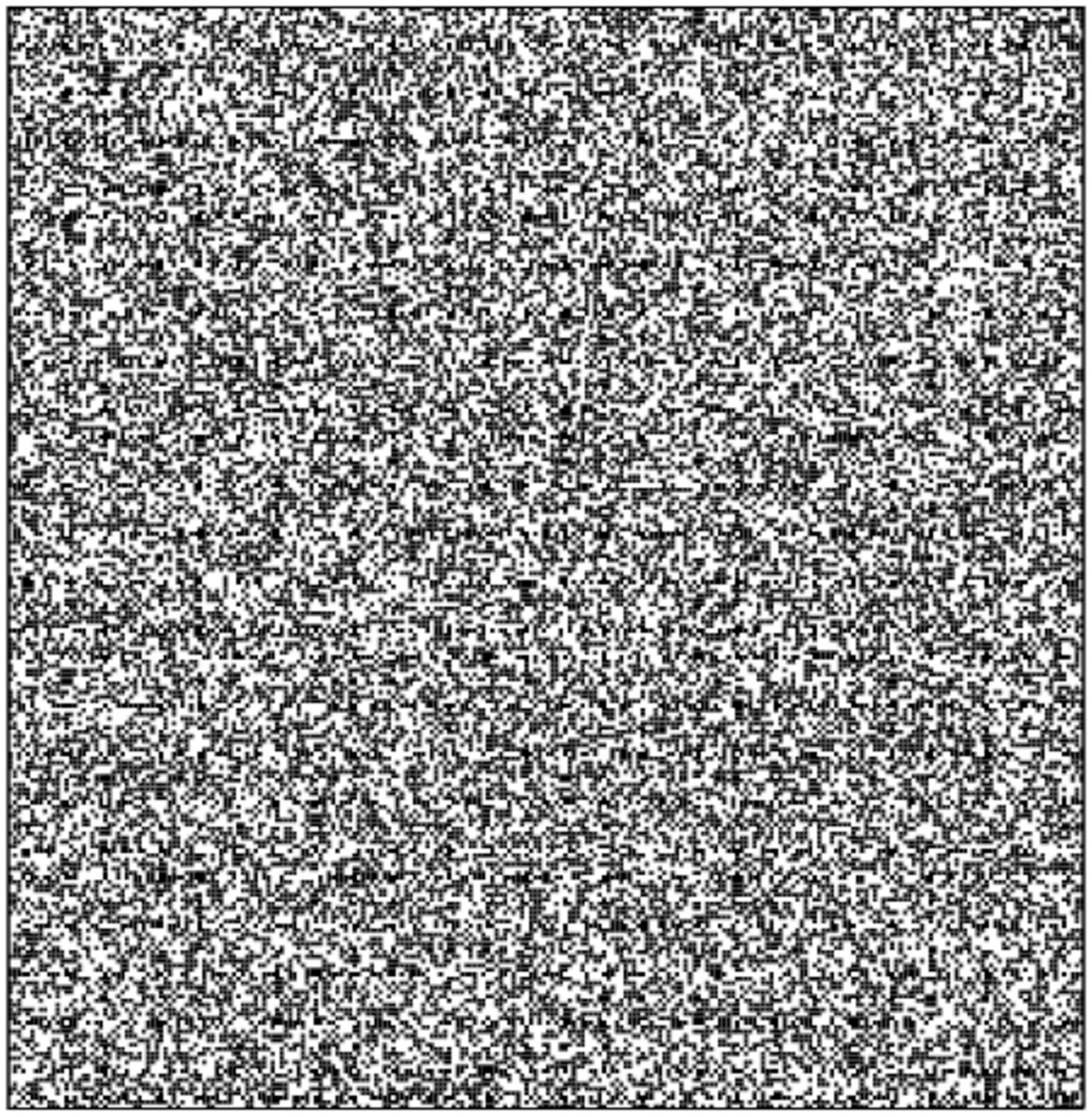}
}
\caption{An equilibrium free-energy landscape with a second-order
phase transition between two phases, a single-valued one with one
minimum, and a bi-valued one with minima related
by parity. Below are three snapshots of a spontaneously broken 
state at $g<g_c$, a critical configuration at  $g=g_c$, and 
a disordered one at $g>g_c$, in the $2d$IM. Black and white dots represent
up and down spins, respectively.}
\label{eq:fig:sketch-transition}
\end{figure}

The static properties of the Ising model (\ref{eq:Ising}) are more
easily analyzed in the experimentally relevant canonical ensemble in
which the system is coupled to a thermal reservoir at temperature
$T$. A second-order continuous phase transition at a finite critical
temperature, $T_c$, separates a high-temperature disordered
paramagnetic phase from a low-temperature ferromagnetic ordered
one. see Fig.~\ref{eq:fig:sketch-transition}. $T_c$ depends on $d$ and
the statistics of the interaction strength. The nature of the two
phases is understood since the development of the Curie-Weiss
mean-field theory while the critical phenomenon has been accurately
described with the help of the renormalization group, that allows for
the computation of all critical exponents~\cite{critical-phenomena}.

\textcolor{black}
{
\subsubsection{The Potts model}}

A natural generalization of the Ising model is the Potts
model~\cite{Wu82} in which the spins $\sigma_i$ take $q$ integer
values from $1$ to $q$. The Hamiltonian is given by
\begin{equation}
{\cal H} = -\sum_{\langle ij\rangle} J_{ij} \delta_{\sigma_i \sigma_j}
\end{equation}
where the sum is over nearest-neighbours on the lattice. Either
uniform, $J_{ij}=J$ (pure case), or bimodal $P(J_{ij})=p
\delta(J_{ij}-J_1) + (1-p) \delta(J_{ij}-J_2)$ (random case)
interactions are usually studied.  In $d=2$ the transition,
discontinuous for $q>4$ and continuous for $q\leq 4$, occurs at
$(e^{\beta_c J_1}-1) (e^{\beta_c J_2}-1) = q$ in the random case with
$p=1/2$. $T_c$ for the pure limit is recovered by setting $J_1=J_2$.

Soap films and general grain growth~\cite{Stavans} are two physical
systems mimicked by Potts models. In $2d$ these physical systems are
made of polygonal-like cells separated by thin walls endowed with line
energy; different `colour' domains correspond to different cells.

\textcolor{black}
{
\subsection{Microscopic dynamics}}

Once the Hamiltonians have been defined, we need to assign updating
rules to the microscopic variables. Classically, the spins do not have
an intrinsic dynamics.  Quite generally, we are interested in the
evolution of physical systems in contact with an environment that provides
thermal agitation and dissipation.  Thus, the evolution is assumed to
be stochastic~\cite{stochastic} and it does not conserve the total
energy. If we want to let the system reach equilibrium at
sufficiently long times we use dynamic rules that satisfy
\textcolor{black}{detailed balance}:
\begin{equation}
W(C\to C') P_{eq}(C) = W(C'\to C) P_{eq}(C') 
\; , 
\end{equation}
with $W(C\to C')$ the transition probability from configuration $C$ to
configuration $C'$ and $P_{eq}(C) =e^{-\beta H(C)}{\sum_{C''} e^{-\beta
H(C'')}}$ the Boltzmann weight.  Finally, one has to decide whether
there are conserved quantities, for instance the global magnetization,
$M=\sum_{i=1}^N s_i$, or a local one, {\it e.g.}
$m_{loc}=s_i+s_j$ with $i$ and $j$ nearest-neighbours. The existence
of conserved quantities puts constraints on the microscopic updates. In
the context of Ising and Potts models two types of dynamics are
common:\\ 
\textcolor{black} {\it Non-conserved order parameter}; there
are no constraints on the stochastic moves and two examples are Monte
Carlo or Glauber dynamics for Ising spins~\cite{Barkema}.\\
\textcolor{black} {\it Conserved order parameter}; it mimics
particle-hole exchanges in lattice gas models obtained from a mapping
of the Ising model and the local magnetization is conserved.
Exchanges of up and down spins are proposed and these are accepted
with a probabilistic law (Kawasaki dynamics~\cite{Barkema}).\\
\textcolor{black}{\it Kinetic constraints} Spin updates are allowed (and 
realized with some stochastic rule) only 
when a chosen constrained is satisfied, typically by neighbor
variables. \\ 

{\it Dynamical systems} are defined by dynamic rules that do not
necessarily satisfy detail balance and have no reference to any
equilibrium energy (or free-energy) that they should minimize.  Up to
what extent the stochastic and deterministic evolution of some
macroscopic systems are similar or even equivalent at some length and
time scales is a relevant question.  Some issues one would like to
understand are under which conditions collective behaviour emerges in
extended dynamical systems with short-range interactions and local
chaotic dynamics; and whether thermodynamic and statistical mechanics
concepts apply in some dynamic regimes, in particular, whether an
energy function and a temperature can be identified in deterministic
dissipative systems.  In spatially extended dynamical systems the
effective noise strength, that may play the role of temperature, is
internally generated rather than imposed by an environment.  Some
lattice models of coupled chaotic maps present dynamic phase
transitions between dynamic phases that can be associated to a
disorder state and a bi-valued dynamically ordered one by using a spin
representation. This is the case of the Miller-Huse coupled map
lattice~\cite{Miller-Huse}, for example, the transition of 
which presents many similarities with critical phenomena.
 
Problems that originate in other sciences, suchlike social studies,
informatics, {\it etc.}  can sometimes be formulated by using stochastic
dynamic rules that do not satisfy detail balance. In some cases,
collective behaviour gives rise to dynamic phase transitions between
phases~\cite{directed-percolation} and by choosing adequately the dynamic rules
one may also encounter a dynamic transition similar to the ones
mentioned  above. This is the case, for instance, of some variants
of {\it voter models}~\cite{voter}.

\textcolor{black}
{
\subsection{The spiral model}}

\begin{figure}
    \centerline{
\hspace{-0.7cm}
   \rotatebox{0}{\resizebox{.45\textwidth}{!}{\includegraphics{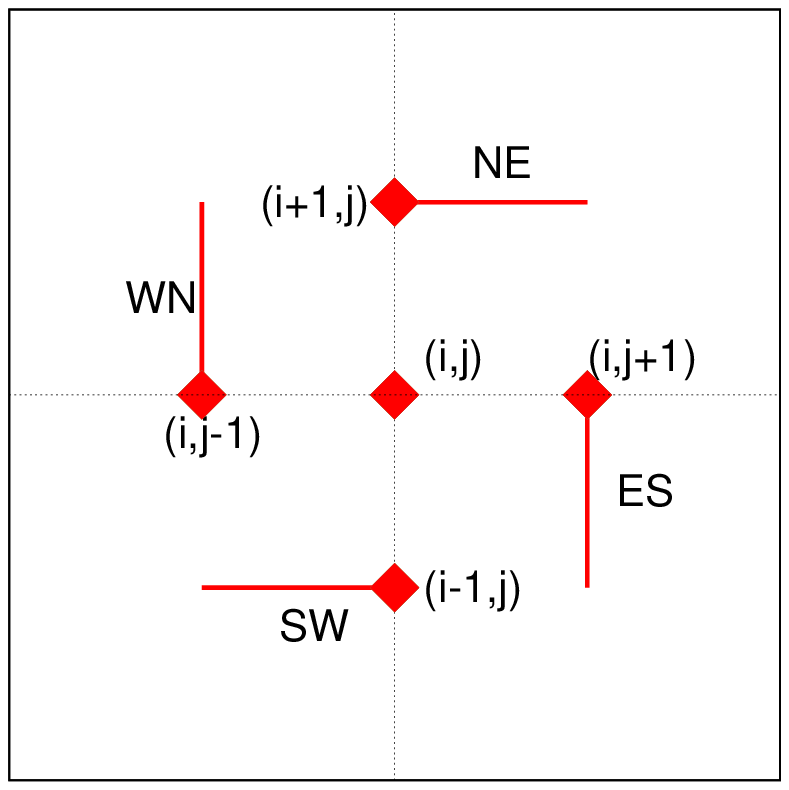}}}
\hspace{-0.3cm}
   \rotatebox{0}{\resizebox{.3\textwidth}{!}{\includegraphics{FIGS/figconfig1.eps}}}
   \rotatebox{0}{\resizebox{.3\textwidth}{!}{\includegraphics{FIGS/figconfig2.eps}}}
}
\vspace{1cm}
    \caption{Left: The neighbouring sites determining the frozen or
free to move character of the center site $(i,j)$.  Centre and right
panel: Configuration of the system at two times ($t=10^5$ and
$t=10^7$) in a quench to $p=0.99$.  Black and white sites are frozen
particles and vacancies, respectively.  Red and green sites are
particles and holes that can be updated. Figure taken from~\cite{Corberi}. }
\label{figsketch}
\end{figure}

Spin kinetically constrained models capture many features of real
glass-forming systems~(see~\cite{kinetically-constrained} for
reviews).  These models display no thermodynamic singularity: their
equilibrium measure is simply the Boltzmann factor of independent
spins and correlations only reflect the hard core constraint.
Bootstrap percolation arguments allowed C. Toninelli {\it et al.} to
show that, when defined on finite dimensional lattices, these models
do not even have a dynamic transition at a particle density that is
less than unity in the thermodynamic limit~\cite{Cristina0}. On Bethe
lattices instead a dynamical transition similar to the one predicted
by the mode coupling theory might occur~\cite{Sellitto}. The quest to
define a finite dimensional kinetically constrained model with a
(discontinuous) transition was positively answered by C. Toninelli
{\it et al} who constructed the so-called spiral model, a finite
dimensional kinetically constrained model with an ideal {\it
  glass-jamm\-ing} dynamic transition at a particle density that is
different from one~\cite{Cristina}.  A binary variable $n_{ij} =0,1$
is defined on the sites $(i,j)$ of an $L\times L$ periodic
bi-dimensional square lattice. $n_{ij}=1$ if a particle occupies the
site and $n_{ij}=0$ otherwise.  One
defines the couples of neighbouring sites (see Fig.~\ref{figsketch}-left):\\
\indent
$(i,j+1)$,$(i+1,j+1)$, north east (NE) couple;\\
\indent
$(i+1,j)$,$(i+1,j-1)$, east south (ES) couple;\\
\indent
$(i,j-1)$,$(i-1,j-1)$, south west (SW) couple;\\
\indent $(i-1,j)$,$(i-1,j+1)$, west north (WN) couple.
\\
Whether the variable $n_{ij}$ may be updated or not depends on the
configuration on these pairs: if the sites belonging to at least two
consecutive couples (namely, NE+ES or ES+SW or WN+NW or WN+NE) are
empty site $(i,j)$ can be
updated (either emptied or filled).  Otherwise it is blocked. Each
site is coupled to a particle reservoir in such a way that particles
can enter or leave the sample from its full volume.  Defining
$M_{ij}=0$ if site $(i,j)$ is frozen and $M_{ij}=1$ otherwise the
updating rule can be expressed in terms of the transition rates
$W(n_{ij}\vert n'_{ij})$ to add or remove a particle as
$W_p(0\vert 1)=M_{ij}\ p$ and 
$W_p(1\vert 0)=M_{ij}\ (1-p)$.
The model can be regarded as a two-level system described by
$H=-\sum _{ij}(2n_{ij}-1)$ at the inverse temperature
$\beta =(1/2)\ln [p/(1-p)]$.
When $p\to 1/2$ the inverse temperature vanishes $\beta \to 0$ whereas
for $p\to 1$ it diverges $\beta \to \infty$.  The Bernoulli measure
implies $\rho=\langle n_{ij}\rangle =p$ for the equilibrium density of
particles.  In equilibrium at zero temperature ($p\to 1$), the lattice
is full while at infinite temperature ($p\to 1/2$) it is half-filled.
The existence of a blocked cluster at $p \geq p_c<1$ was shown by
studying the critical $p$ above which an equilibrium configuration
cannot be emptied and this coincides with the threshold for $2d$
directed percolation ($p_c\simeq 0.705$).
At the
transition an infinite cluster of blocked particles exists and the
density of the frozen cluster is discontinuous at $p_c$. Many 
aspects of super-cooled liquids
slowing down are reproduced close (but below) $p_c$~\cite{Cristina}.
In Sect.~7
we shall summarize results in~\cite{Corberi}
where we studied, in particular,  the dynamics following
a quench from $p_0<p_c$ to $p>p_c$.

\textcolor{black}
{
\section{Geometric description of phase transitions}}

\textcolor{black}{Geometric domains} are ensembles of connected
nearest-neighbour sites with spins pointing in the same
direction. Their area is the number of spins belonging to the domain.
By lumping together with a certain temperature-dependent probability
neighboring spins in the same spin state, spin models can be mapped
onto percolation theory. The resulting
\textcolor{black}{Fortuin-Kasteleyn spin clusters}~\cite{FK} built using
$p_{ij}=1-e^{-\beta J_{ij}}$ percolate at the critical temperature,
and their percolation exponents coincide with the thermal ones. In
this way, a purely geometrical description of the phase transition 
is achieved.  In the rest of this Section we sum up the
definitions of the main geometric objects used in such a static
description of critical phenomena.

\textcolor{black}
{
\subsection{Definitions}}

A number of \textcolor{black}{linear dimensions} of a cluster can be
defined.  The \textcolor{black}{radius of gyration} is 
\begin{equation}
R_g^2=s^{-1} \sum_{i=1}^s |\vec r_i-\vec r_{cm}|^2
\end{equation}
 with the centre of mass position given by
$\vec r_{cm}=s^{-1} \sum_{i=1}\vec r_i$.

Take the cluster site with the largest (smallest) $y$ component and
among these the one with the largest (smallest) $x$ component. These
sites are the two `end-points' of the cluster. The
\textcolor{black}{end-to-end} length of the cluster, $\ell$,  is the Euclidean
distance between these two points.

There are also several ways of defining the \textcolor{black}{surface}
of a cluster~\cite{percolation}.  Different definitions are relevant
to different applications that have to do, for example, with
the adsorption of particles on the surface.  Without entering into all the
zoology let us recall some of these definitions.

\begin{figure}
\begin{center}
\includegraphics[width=5.5cm]{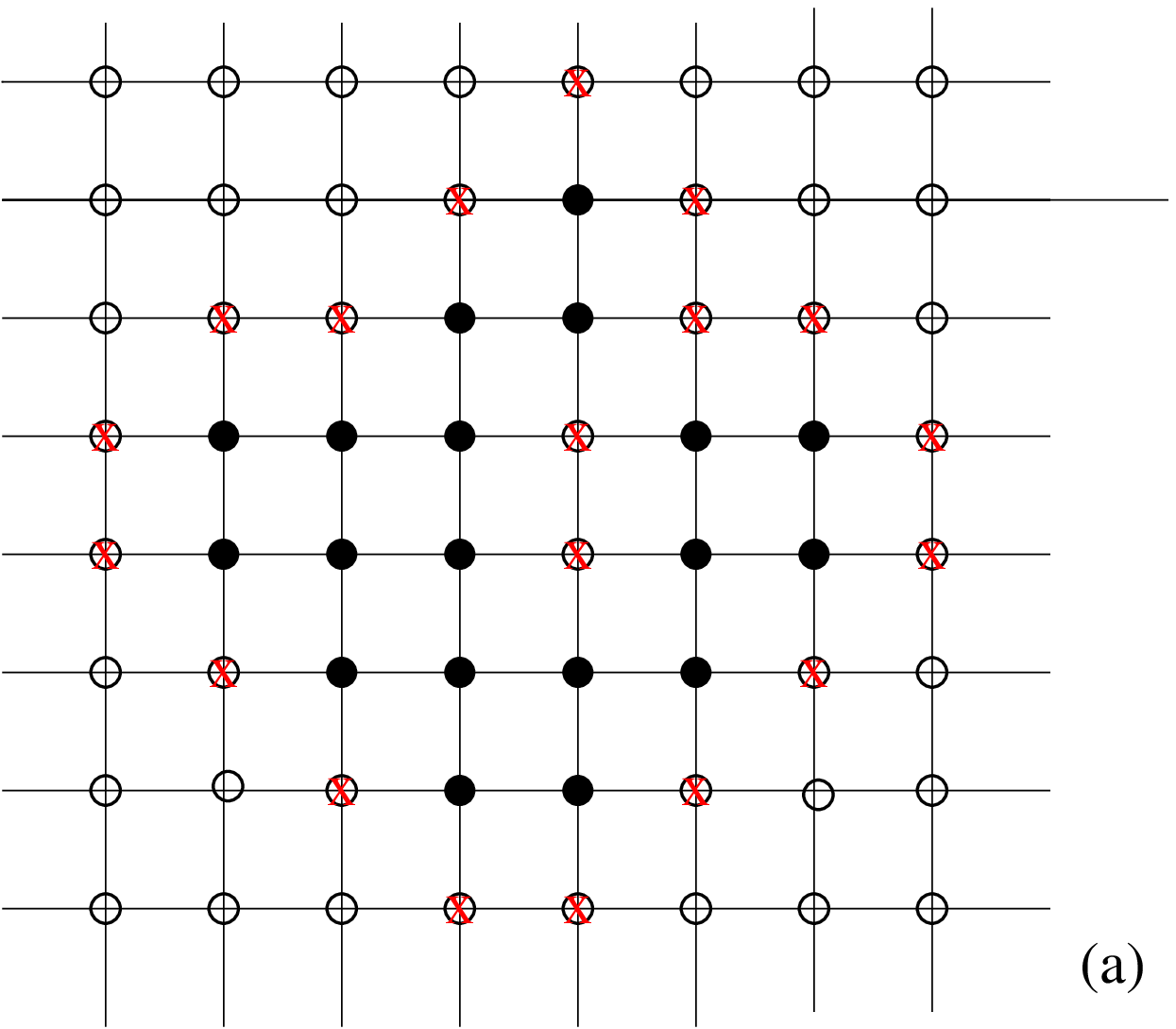}
\includegraphics[width=5.5cm]{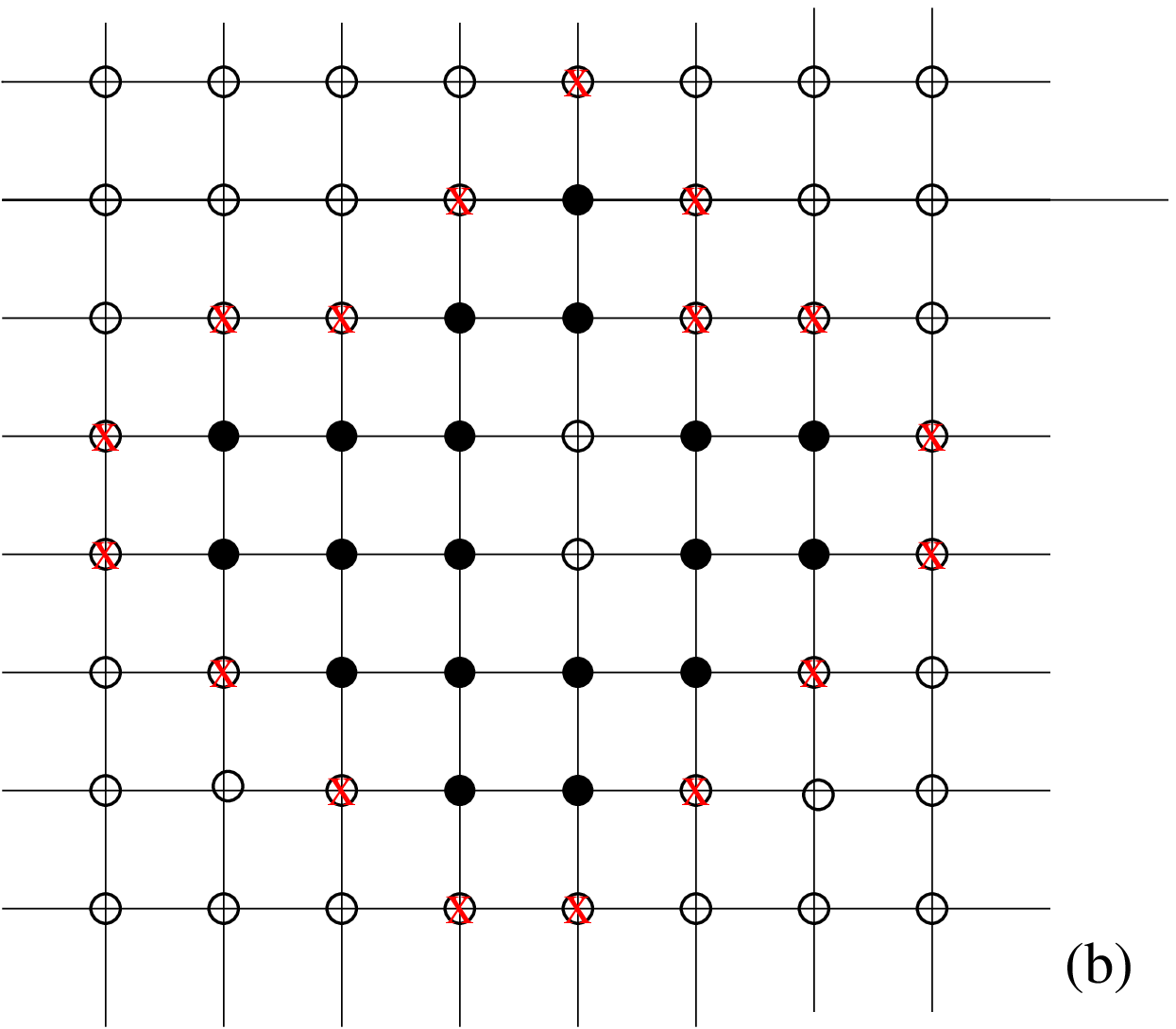}
\end{center}
\caption{Square bi-dimensional lattice. Filled (empty) circles
represent occupied (empty) sites. Red crosses indicate
external border of the cluster of filled sites including (a), or nor
(b), vacant sites that are connected to the exterior by a next nearest
neighbour (a `diagonal').}
\label{fig:sketch-external-border}
\end{figure}

The \textcolor{black}{external border} of a 
cluster of occupied sites is the set of
vacant sites that are nearest-neighbours to sites on the cluster and
are connected to the exterior by (a) nearest {\it and} next-to-nearest
neighbours or (b) just next nearest neighbours.
Figure~\ref{fig:sketch-external-border}~(a) and (b) show the external
borders of a cluster of occupied sites thus defined. Note that in (a)
there are two `inner' sites that belong to the external border and
that are connected to the outside by a narrow neck of width
$\sqrt{2}a$ with $a$ the lattice spacing while in (b) these two sites
do not belong to the border and the `fjord' has been excluded.

\begin{figure}
\begin{center}
\includegraphics[width=5cm]{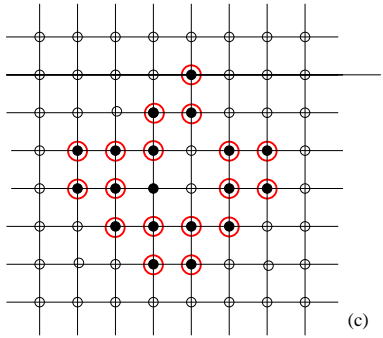}
\end{center}
\caption{Same cluster as in Fig.~\ref{fig:sketch-external-border}.
Encircled are the sites on the cluster that belong to its hull.
The hull-enclosed area is everything that lies within the hull.}
\label{fig:sketch-hull}
\end{figure}

The \textcolor{black}{hull} is the envelop of a cluster, meaning the
ensemble of cluster sites obtained by using a biased walk that
encircles the cluster on its left and on its right linking the lowest
lying site to the left and the highest lying site to the right, see
Fig.~\ref{fig:sketch-hull}. The hull of a cluster of occupied (vacant)
sites lies on occupied (vacant) sites. 

The \textcolor{black}{hull-enclosed} area is the total area within the hull 
(including the sites on the hull). It therefore ignores the type of 
site (occupied or vacant, spin up or spin down) that lies within and 
counts them all on equal footing.

\textcolor{black}
{
\subsection{Number densities}}

We call $n(s)$ the number of objects (hulls, clusters, {\it etc.})
with $s$ sites per unit number
of lattice sites. Scaling theory, some exactly solvable cases ({\it e.g.}
the Bethe lattice), renormalization group arguments~\cite{percolation}, 
Coulomb gas mappings, 
conformal field theory calculations, stochastic Loewner evolution 
techniques~\cite{Duplantier,Vanderzande-Stella} and numerical experiments~\cite{Janke}
suggest
\begin{equation}
n(s) \sim s^{-\tau} \ f[\theta s^\sigma] \qquad \qquad 
\mbox{for large} \;\; s
\; , 
\label{eq:ns-1}
\end{equation}
with $\tau$ and $\sigma$ two control parameter and lattice-independent
exponents that do depend on the space dimensionality and $\theta$
measuring the distance from criticality, say $\theta=|g-g_c|$ with $g$
the control parameter.  
The function $f(z)$ approaches a
constant for $|z|\ll 1$ and it falls-off rapidly for $|z|\gg 1$; it
thus provides a cut-off with a {\it unique} cross-over size
$s_\xi\equiv |g-g_c|^{-1/\sigma}$ such that clusters with $s<s_\xi$
are effectively critical while those with $s>s_\xi$ are not.  
At criticality ($g=g_c$) the factor $f[\theta
s^\sigma]=f[0]$ that suppresses large objects is absent and clusters of
all sizes are present.

The exponents $\sigma$ and  $\tau$
determine all critical exponents of percolation through scaling
relations.  As regards thermal
phase transitions, once the relevant Fortuin-Kasteleyn clusters are
constructed and analysed their two independent exponents $\sigma$ and
$\tau$ determine the entire set of thermal critical exponents (see,
however,~\cite{Alberto}).

\textcolor{black}
{
\subsection{Fractal exponents}}

Geometric objects at criticality have a fractal behaviour, that is to
say~\cite{percolation}
\begin{equation}
s \propto \ell^{D}
\; . 
\end{equation}
$s$ is the mass of the object (cluster area, its external border, its
hull), $\ell$ is the linear size of the cluster, say its radius of
gyration, and $D$ is the fractal dimension.  The fractal dimension $D$
is related to the exponent $\tau$ in eq.~(\ref{eq:ns-1}) via
$\tau=d/D+1$ where $d$ is the space dimension.

The fractal exponents depend strongly on the precise definition of the
geometric object. For instance, Saleur and Duplantier showed that $2d$
critical percolation clusters have $D_E=4/3$ and $D_H=7/4$ where $H$
stands for hull and $E$ for external border connected to the exterior
by nearest-neighbour vacant sites only~\cite{Saleur-Duplantier} (case (b) in
Fig.~\ref{fig:sketch-external-border}).

In~\cite{Sicilia-PRE} we studied the relation between areas and perimeters
\begin{equation}
A\propto p^\alpha
\end{equation}
with $p$ defined as the number of broken bonds on the external border
or, equivalently, with a small variant of the Grossman-Aharony
construction~\cite{percolation} including fjords.  Note that if
$A\propto \ell^{D_A}$ and $p\propto \ell^{D_p}$ with $\ell$ the linear
size of the object, say the radius of gyration, then $\alpha=D_A/D_p$.

\textcolor{black}{\section{Coarsening}}

Take a system in equilibrium in the symmetric (positive $g-g_c$, high
temperature) phase and quench it into the symmetry breaking (negative
$g-g_c$, low temperature) phase through a second order phase
transition. For concreteness, we focus on problems with two
equilibrium states related by $Z_2$ symmetry. Once set into the
ordered phase the system {\it locally} selects one (among the
two possible) equilibrium configurations. However, different `states' are
picked up at different locations and \textcolor{black}{topological
  defects} in the form of \textcolor{black}{domain walls} are
created. In the course of time the patches of ordered regions tend to
grow while the density of topological defects diminishes. In an
infinite system this \textcolor{black}{coarsening process} goes on
forever.

An example of the above process is given by an Ising magnet. An
initial condition in equilibrium at, for instance, infinite
temperature is just a random configuration in which each spin takes
one of its possible values with probability $1/2$. After a quench
below the critical temperature, $T<T_c$, the ferromagnetic
interactions tend to align the neighbouring spins in `parallel'
direction (same value of the spin) and in the course of time domains
of the ordered phases form and grow.  At any finite time the
configuration is such that both types of domain exist.  Under more
careful examination one reckons that there are some spins reversed
within the domains. These `errors' are due to thermal fluctuations and
are responsible of the fact that the magnetization of a given
configuration within the domains is smaller than one and close to the
equilibrium value at the working temperature (apart from fluctuations
due to the finite size of the domains). At each instant there are as
many spins of each type (up to fluctuating time-dependent corrections
that vanish in the infinite size limit).  As time
passes the typical size of the domains increases in a way that we
shall discuss below.

\begin{figure}
\begin{center}
\includegraphics[width=5cm]{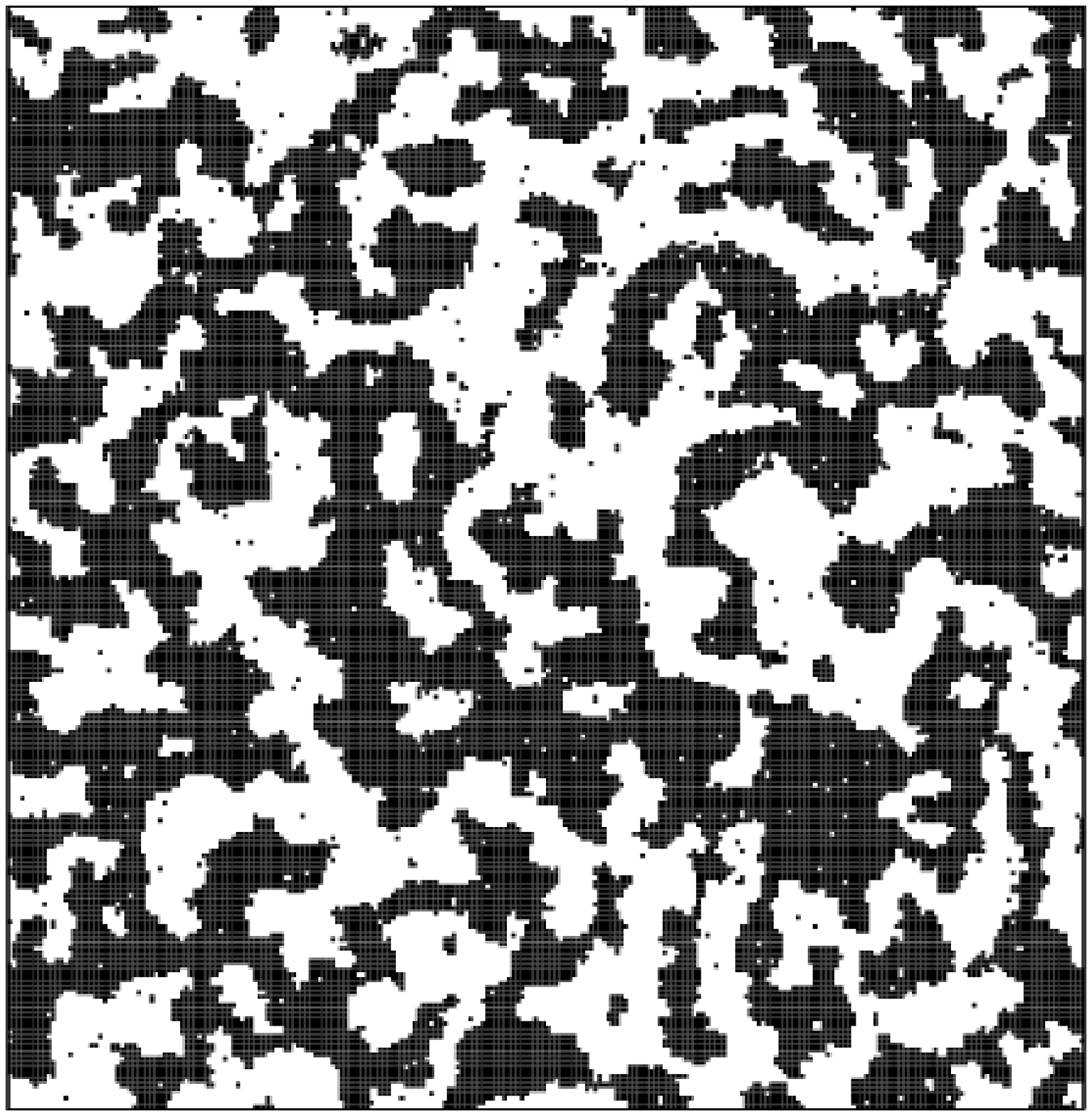}
\includegraphics[width=5cm]{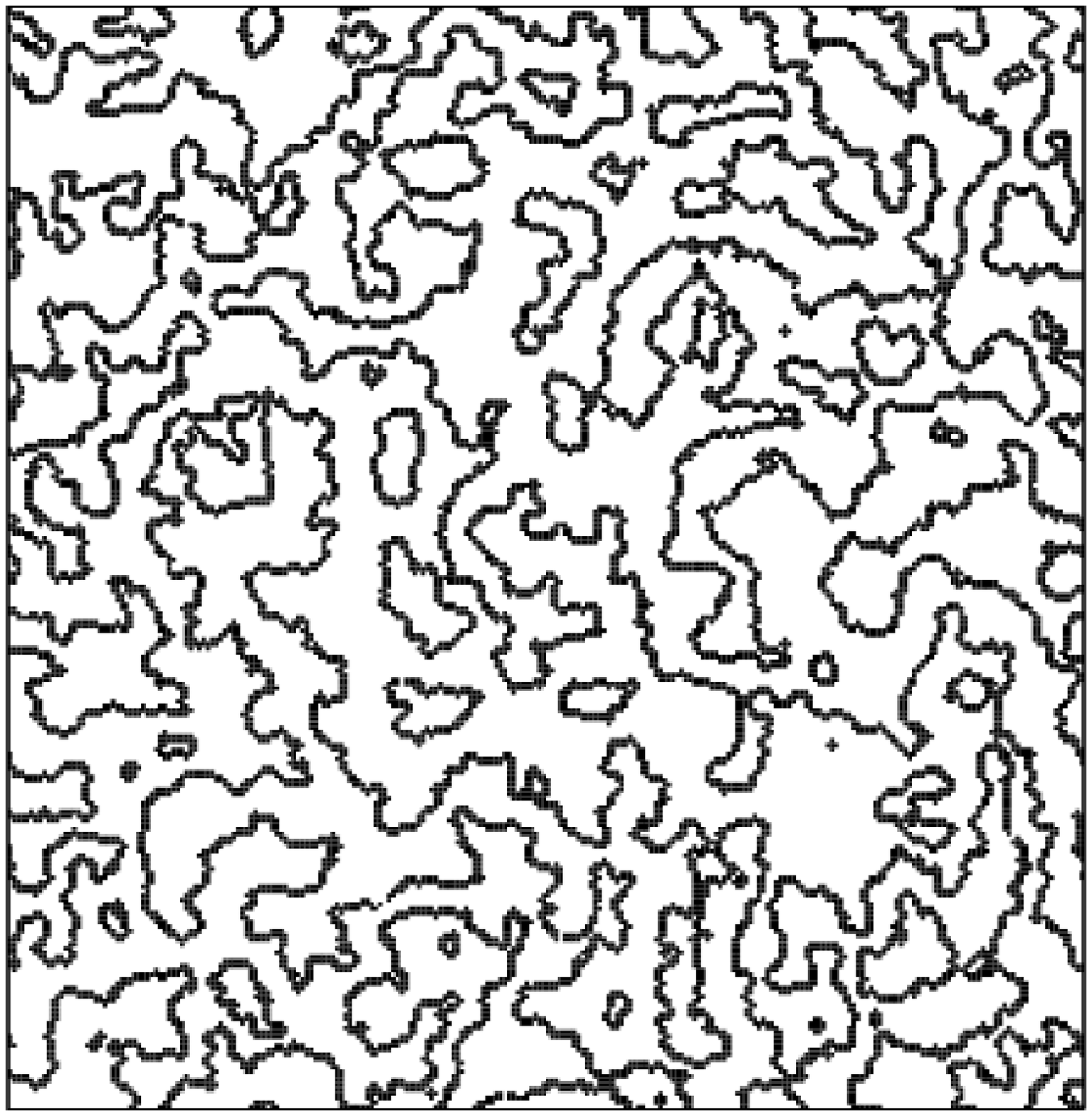}
\end{center}
\caption{Snapshot of the $2d$ Ising model at a number of Monte
  Carlo steps after a quench from infinite to a sub-critical temperature.
  Left: the up and down spins on the square lattice are represented
  with black and white sites. Right: the domain walls are shown in
  black. Figure taken from~\cite{Sicilia-PRE}.}
\label{fig:domain-growth-photo}
\end{figure}

Coarsening from an initial condition that is not correlated with the
equilibrium state and with no bias field does not take the system to
equilibrium in finite times with respect to a function of the system's
linear size, $L$. More explicitly, if the growth law is a power law
[see eq.~(\ref{eq:zd})] one needs times of the order of $L^{z_d}$ to
grow a domain of the size of the system. This gives a rough estimate of
the time needed to take the system to one of the two equilibrium
states.  For any shorter time, domains of the two types exist and the
system is \textcolor{black}{ out of equilibrium}.

We thus wish to distinguish the \textcolor{black}{ relaxation time},
$t_r$, defined as the time needed for a given initial condition to
reach equilibrium with the environment, from the \textcolor{black}{
decorrelation time}, $t_d$, defined as the time needed for a given
configuration to decorrelate from itself.  At $T<T_c$ the relaxation
time of {\it any} initial condition that is not correlated with the
equilibrium state diverges with the linear size of the system.  The
self-correlation of such an initial state evolving at $T<T_c$ decays
as a power law and although one cannot associate to it a decay time as
one does to an exponential, one can still define a characteristic time
that turns out to be related to the age of the system.

In contrast, the relaxation time of an {\it equilibrium} magnetized
configuration at temperature $T$ vanishes since the system is already
in equilibrium while the decorrelation time is finite and given by $t_d
\sim |T-T_c|^{-\nu z_{eq}}$.

The lesson to learn from this comparison is that the relaxation time
and the decorrelation time not only depend on the working temperature
but they also depend strongly on the initial condition. Moreover, 
the relaxation time depends
on $(N,T)$ while the decorrelation time depends on $(T,t_w)$. For a 
random initial condition one has 
\begin{eqnarray}
t_r \simeq \left\{
\begin{array}{ll}
\mbox{finite} \qquad\qquad & T>T_c
\; , 
\nonumber\\
|T-T_c|^{-\nu z_{eq}} \qquad\qquad & T\stackrel{>}{\sim} T_c
\; , 
\nonumber\\
L^{z_d} \qquad\qquad & T<T_c
\; . 
\end{array}
\right.
\end{eqnarray}

\textcolor{black}
{
\subsection{Critical coarsening}}

Right after a quench to the critical point the system starts to evolve
out of equilibrium towards a target equilibrium configuration of the
kind shown in the central snapshot in
Fig.~\ref{eq:fig:sketch-transition}. The non-equilibrium relaxation
shows coarsening features with the growth of ordered structures with a
typical linear length given by $R_c(t) \simeq t^{1/z_{eq}}$, and
features of equilibrium dynamics, as the fact that $z_{eq}$ is the
equilibrium dynamics exponent. Interestingly enough, the early
relaxation exhibits universal scaling properties characterized by
usual static exponents as well as the dynamic one. This and other
details of the critical equilibrium and out of equilibrium dynamics
can be computed with renormalization group
techniques~\cite{Beate,short-time-dyn}.

\textcolor{black}
{
\subsection{Dynamic scaling hypothesis}}

The \textcolor{black}{dynamic scaling hypothesis} states that at late
times and in the scaling limit~\cite{review-coarsening}
\begin{equation}
r\gg \xi(g) \; , \qquad R(g,t) \gg \xi(g) \; , \qquad r/R(g,t) \;\;
\mbox{arbitrary} \; ,
\label{eq:scaling-limit}
\end{equation}
where $r$ is the distance between two points in the sample, $r\equiv
|\vec x -\vec x'|$, and $\xi(g)$ is the equilibrium correlation length
that depends on all parameters ($T$ and possibly others) collected in
$g$, there exists a \textcolor{black}{ single characteristic length},
$R(g,t)$, such that the domain structure is, in statistical sense,
independent of time when lengths are scaled by $R(g,t)$. Time, denoted
by $t$, is typically measured from the instant when the critical point
is crossed.  In the following we ease the notation and write only the
time-dependence in $R$. This hypothesis has been proved analytically
in very simple models only, such as the one dimensional Ising chain
with Glauber dynamics or the Langevin dynamics of the $d$-dimensional
$O(N)$ model in the large $N$ limit. In the vast majority of
coarsening systems the dynamic scaling hypothesis applies. Still, a
few counter-examples where two lengths grow in competition are also
known~\cite{review-coarsening}.

\begin{figure}[h]
\centerline{
\psfrag{r}{$r/R(t)$} \psfrag{cr}{$C(r/R(t))$} \psfrag{t}{$t$}
\psfrag{r2}{$R^2(t)$} \psfrag{tt}{$t$}
\includegraphics[width=9cm]{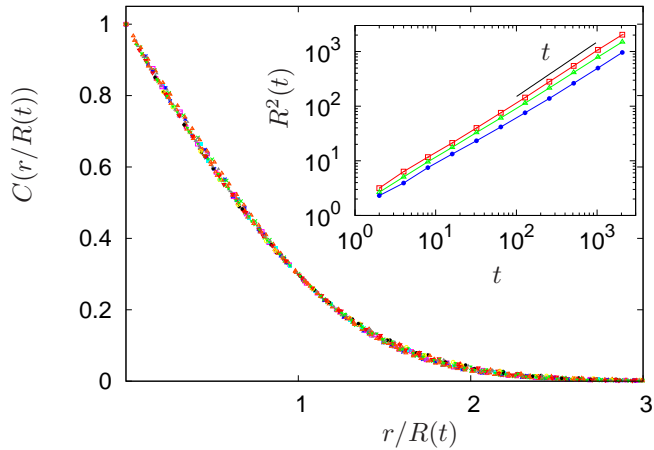}
}
\caption{Check of dynamic (super-)scaling in the space-time
correlation of the Potts model quenched from $T_0\to\infty$ to $T_f=
T_c(q)/2$.  Data taken at several times (from $t=2^4$ MCs to $2^{11}$
MCs) for $q=2$, 3 and 8, with and without weak disorder are
shown. Distance is rescaled by the $R(t)$ obtained from $C(R,t)=0.3$.
Inset: $R^2(t)$ against $t$ in a double logarithmic scale for $q=2$, 3
and 8 (from top to bottom).  The characteristic length, related to the
average domain radius, depends weakly on $q$ and $T$ for the pure
model (through the pre-factor). Within the time window explored there
are still some small deviations from the $R(t) \simeq t^{1/2}$
expected law in the case $q=8$.  When weak disorder is introduced (not
shown), the growth rate is greatly reduced and deviates from the
linear behavior, see Sect.~5.2.  Figure
taken from~\cite{Loureiro}.  }
\label{fig:scaling-Crt}
\end{figure}

In practice, the dynamic scaling hypothesis implies that 
the real-space correlation function should 
behave as
\begin{equation}
C(r,t) \equiv \langle s_i s_j \rangle_{|\vec r_i-\vec r_j|=r} \simeq  F[r/R(t)]
\; . 
\end{equation}
See Fig.~\ref{fig:scaling-Crt} for a test of the scaling hypothesis in
the MC dynamics of the $2d$ Ising and Potts model (see the caption for
details).  In order to fully characterise the correlation functions
one then has to determine the typical growing length, $R$, and the
scaling functions, $f_c$, $F$, {\it etc.}

\textcolor{black}{
\subsubsection{The growing length}
}

It turns out that $R$ can be
determined with semi-analytic arguments and the predictions are well
verified numerically -- at least for clean system. 
In \textcolor{black}{pure and isotropic} systems the growth law is 
\begin{equation}
R(t) = \lambda \ t^{1/z_d}
\label{eq:zd}
\end{equation}
with $z_d$ the \textcolor{black}{dynamic exponent}
(see~\cite{review-coarsening}) and $\lambda$ a material/model dependent
prefactor that weakly depends on temperature and other parameters.  
In curvature driven Ising or Potts
cases with non-conserved order parameter the domain walls have finite width and
$z_d=2$. For systems with continuous variables such as rotors or XY
models and non-conserved order-parameter dynamics, a number of computer simulations
have shown that domain walls are thicker and $z_d=4$.  The effects of
temperature enter only in the parameter $\lambda$ and, for clean
systems of Ising type, growth is slowed down by 
temperature since thermal fluctuation tend to roughen the interfaces
thus opposing the curvature driven mechanism.
Let us list some special cases below and
sketch how $z_d$ can be estimated. 

\noindent{\it \textcolor{black}{
Clean one dimensional cases with non-conserved order parameter}
}

In one dimension, a space-time graph allows one to view coarsening as
the diffusion and annihilation upon collision of point-like particles
that represent the domain walls. In the Glauber Ising chain with
non-conserved dynamics one finds that the typical domain length grows
as $t^{1/2}$ while in the continuous case the growth is only
logarithmic, $\ln t$.

\noindent{\it \textcolor{black}{
Non-conserved curvature driven dynamics ($d>2$)}
}

The time-dependent Ginzburg-Landau model allows us to gain some
insight on the mechanism driving the domain growth and the direct
computation of the averaged domain length. This is a stochastic partial
differential equation on a coarse-grained order parameter field with a
deterministic force that is phenomenologically proposed to derive from
a Ginzburg-Landau type free-energy.  In clean systems temperature does
not play a very important role in the domain-growth process, it just
adds some thermal fluctuations within the domains, as long as it is
smaller than $T_{c}$. In dirty cases instead temperature triggers
thermal activation.

We focus first on zero temperature clean cases with Ising-like
symmetry. At $T=0$ the GL equation is just a gradient descent in a
(free-)energy landscape, $F$.  Two terms contribute to $F$: an elastic
energy $(\nabla \phi )^{2}$ which is minimized by flat walls if
present and a bulk-energy term that is minimized by constant field
configurations, say $\phi=\pm \phi_0$ in an Ising-like case.  If the
walls are sharp enough, that is to say their width remains 
finite when the distance between them diverges, 
interface-interface interactions can be neglected and
the minimization process implies that regions of constant field grow
and get separated by flatter and flatter walls. Within this scenario
one can easily derive the \textcolor{black}{Allen-Cahn
equation}~\cite{Allen-Cahn} that states that the local wall velocity 
is proportional to the local geodesic curvature and 
is normal to the wall pointing in the direction of reducing curvature:
\begin{equation}
\vec v\equiv \partial_t \hat n|_\phi =- \vec \nabla \cdot
  \hat n \ \hat n \equiv -\frac{\lambda}{2\pi} \kappa
\hat n 
\; , 
\label{eq:velocity-curvature}
\end{equation}
in all $d$.
This equation yields an intuition on the typical growth law in such
processes. Take a spherical wall in any dimension. The local curvature
is $\kappa = (d-1)/R$ where $R$ is the radius of the sphere within the
wall.  Equation~(\ref{eq:velocity-curvature}) is recast as $dR/dt = -
\lambda (d-1)/R$ that implies $R^2(t) = R^2(0) - 2\lambda (d-1)t$ and
$R$ \textcolor{black}{decreases} as $t^{1/2}$. This calculation implies
that all hull-enclosed areas decrease in time with the same law,
independently of their own size and of all others. Interestingly
enough, similar results are obtained for the area loss of a
square domain embedded in a sea of the opposite sign using 
Monte Carlo or Glauber dynamics at zero temperature for a system
defined on a square lattice~\cite{Kandel}.

 The above results do not imply that all domains shrink as well since
some gain size from the disappearance of inner objects. Temperature
effects are simply taken into account by a material and $T$-dependent
proportionality constant $\lambda$ in the Allen-Cahn equation.

In $d=2$ the time dependence of the area contained within
any finite hull on a flat surface is derived 
by integrating the velocity around the envelope and using the Gauss-Bonnet
theorem:
\begin{equation}
\frac{dA}{dt}= 
\oint \vec v \wedge \vec \ell
= \oint v dl = -\frac{\lambda}{2\pi}\oint \kappa dl
=-\lambda\left(1-\frac{1}{2\pi}\sum_i \theta_i\right),
\end{equation}
where $\theta_i$ are the turning angles of the tangent vector to the
surface at the $n$ possible vertexes or triple junctions between
domains of different colour (we are now generalizing the discussion to
models with multi-valued equilibrium states suchlike 
the Potts model with $q\geq 2$).  In  the
Ising $q=2$ model, $\sum_i \theta_i=0$ since there are no such
vertexes and we obtain $dA/dt=-\lambda$ for all hull-enclosed areas,
irrespective of their size. If, instead, like in soap froths, there is 
a finite number of such vertexes with an angle of $2\pi/3$, that is,
$\theta_i=\pi/3, \forall i$ (for highly anisotropic systems, as the
Potts model, the angles are different from $2\pi/3$).
The above equation thus reduces to the von Neumann
law~\cite{Neumann52} for the hull-enclosed area $A_n$ with
a hull with $n$-turning angles:
\begin{equation}
\frac{dA_n}{dt}=\frac{\lambda}{6}(n-6)
\label{eq:Neumann}
\end{equation}
(in the Ising case $n=0$).
Whether a cell grows, shrinks or remains with constant area depends
on its number of sides being, respectively, larger than, smaller than 
or equal to 6.

\noindent{\it \textcolor{black}{
Conserved order parameter: bulk diffusion}
}

A different type of dynamics occurs in the case of phase separation
(a water and oil mixture ignoring hydrodynamic interactions or a
binary alloy).  In this case, the material is locally conserved, {\it
  i.e.}  water does not transform into oil but they just separate. The
main mechanism for the evolution is diffusion of material through the
bulk of the opposite phase.  After some discussion, it was
established, as late as in the early 90s, that for scalar systems with 
\textcolor{black}{conserved order parameter} $z_d=3$~(see~\cite{Huse,Barkema}).


\noindent{\it \textcolor{black}{
Role of disorder: thermal activation}}

The situation becomes less clear when there is quenched disorder
in the form of non-magnetic impurities in a magnetic sample, lattice
dislocations, residual stress, {\it etc}.  Qualitatively, the dynamics
are expected to be slower than in the pure cases since disorder pins
the interfaces. In general, based on an argument due to Larkin (and in
different form to Imry-Ma) one expects that in $d<4$ the late epochs
and large scale evolution be no longer curvature driven but controlled
by disorder. 

The argument to estimate the growth law in dirty systems is
the following.  Take a system in one equilibrium state with a domain
of linear size $R$ of the opposite equilibrium state within it. This
configuration could be the one of an excited state with respect to the
fully ordered one with absolute minimum free-energy. Call $\Delta F(R)$ the
free-energy barrier between the excited and equilibrium
states. The thermal activation argument
yields the activation time scale for
the decay of the excited state ({\it i.e.}  erasing the domain wall) 
\begin{equation}
t_A \sim \tau_0 \  e^{\Delta F(R)/(k_B T)}
\; . 
\label{eq:Arrhenius}
\end{equation} 
For a barrier growing as a power of $R$,  
$
\Delta F(R) \sim \Upsilon(T,J) R^\psi
$ (where $J$ represents the disorder) 
 one inverts (\ref{eq:Arrhenius}) to find the 
linear size of the domains still existing at time $t$, 
that is to say, the growth law~\cite{Huse-Henley} 
\begin{equation}
R(t) \sim \left( \frac{k_B T}{\Upsilon(T,J)} \; 
\ln \frac{t}{\tau_0} \right)^{1/\psi}
\; . 
\label{eq:activated-scaling} 
\end{equation}
All smaller fluctuation would have disappeared at $t$ while typically
one would find objects of this size. The exponent $\psi$ is expected
to depend on the dimensionality of space but not on temperature. In
`normal' systems $\psi$ should be $d-1$ -- the surface of the
domain. The pre-factor
$\Upsilon$ is expected to be weakly temperature dependent.

To extend this result to the actual out of equilibrium coarsening
situation
one assumes that the same argument applies out of equilibrium to the
re-conformations of a portion of any domain wall or interface where $R$
is the observation scale.

However, not even for the (relatively easy) random ferromagnet there
is consensus on the actual growth law~\cite{Greg}. We shall discuss a
possible way out in Sect~5.1.  In the case of
spin-glasses, if the mean-field picture with a large number of
equilibrium states were realized in finite dimensional models, the
dynamics would be one in which all these states grow in competition.
If, instead, the phenomenological droplet model applied, there would
be two types of domains growing, $R(t) \sim (\ln t)^{1/\psi} $ and the dimension
of the bulk of these domains should be compact but their surface
should be rough with fractal dimension $d_s>d-1$~\cite{Fisher-Huse}.

\textcolor{black}{\subsubsection{Scaling functions}}

The scaling functions are harder to obtain. Indeed, there is no
systematic method to derive them and all approximations dealt with are
not accurate enough.  For a much more detailed discussion of these
methods see the review articles in~\cite{review-coarsening}.  Still, 
the super-universality property seems to be correct as we discuss in
Sect.~5.2.

\textcolor{black}
{
\section{Dynamics of (weakly) random systems}}
\label{sec:disorder}

In this Section I explain that the existence of a static typical
length yields a natural crossover between the clean growth law, say
$R\simeq [\lambda(T) t]^{1/2}$, and the activation-ruled one,
$R \simeq [k_B T/\Upsilon \ln t]^{1/\psi}$, and how these two
regimes might not be sufficiently separated in numerical and experimental
measurements, being easily confused with a disorder and temperature
dependent power~\cite{Sebastian}.

I also discuss the super-universality hypothesis~\cite{Fisher-Huse}
and list some of its checks.

\textcolor{black}
{
\subsection{Crossover in the  growing length}}
\label{sec:disorder-R}

Numerical simulations of dirty systems tend to indicate that the
growing length is a power law with a disorder and $T$-dependent
exponent. This can be due to the effect of a disorder and
$T$-dependent cross-over length, as explained in~\cite{Sebastian}. 
For concreteness, let us assume that
the width of the quenched random distribution is characterised by a
parameter $J$ and that the cross-over length depends on
$J/T$.  We call the latter $L_T$ by absorbing the $J$
dependence in $T$. The proposal is
that below $L_T$ the growth process is as in the clean limit while
above $L_T$ quenched disorder is felt and the dynamics are thermally
activated above barriers that are usually taken to grow as a power of
the size leading to eq.~(\ref{eq:activated-scaling}): 
\begin{eqnarray}
R(t) &\sim& 
\left\{
\begin{array}{ll}
[\lambda(T) t]^{1/z_d} \qquad \;\;\;\;\; 
& \mbox{for} \qquad R(t) \ll L_T \; , 
\\
\left[k_B T/\Upsilon(T) \ln t\right]^{1/\psi} \qquad 
& \mbox{for} \qquad R(t) \gg L_T \; .
\end{array}
\right.
\end{eqnarray}
These growth-laws can be first inverted to get the time 
needed to grow a given length and then combined into a single expression 
that interpolates between the two regimes:
\begin{equation}
t(R) \sim e^{(R/L_T)^\psi} R^{z_d}
\end{equation}
where the relevant $T$-dependent length-scale $L_T$ has been
introduced. Now, by simply setting $t(R) \sim R^{\overline z(T)}$ one
finds 
\begin{equation}
\overline z(T) -z_d \simeq z/L_T^\psi 
\qquad\qquad \mbox{for times such that} \;\;\;
t^{\psi/z_d} \ln t \simeq \mbox{ct}
\; . 
\end{equation} 
Similarly, by equating $t(R) \sim
\exp(R^{\overline\psi(T)}/k_B T)$ one finds that $\overline \psi(T)$ is a
decreasing function of $T$ approaching $\psi$ at high $T$.

\textcolor{black}
{
\subsection{Super-universality}}
\label{sec:disorder-superscaling}

In its first version
the super-universality hypothesis states that in cases in which
temperature and quenched disorder are `irrelevant' in the sense that
they do not modify the nature of the low-temperature phase ({\it e.g.}
it remains ferromagnetic in the case of ferromagnetic Ising models
with quenched random interactions) the scaling functions should not be
modified~\cite{Fisher-Huse}.  Only the growing length changes from
the, say, curvature driven $t^{1/2}$ law to an asymptotically slower
law due to domain wall pinning by impurities.  Tests of the disorder
independence of the scaling functions of several correlations in
quenches from $T>T_c$ into the ordered phase in the $3d$
RFIM~\cite{RFIM} and the $2d$ RBIM~\cite{RBIM,Sicilia-EPL} (perhaps
excluding the possibility of having zero bonds, see the discussion in
Henkel \& Pleimling) give support to this hypothesis.

We recently investigated two aspects of the super-universality
hypothesis.  On the one hand, we focused on the dependence on the
initial condition of the space-time correlation scaling functions in
the low temperature phase. With this aim, we studied the dynamics of
the Potts model with different values of $q$ after quenches from
$T_0\to\infty$ and $T_0=T_c$~\cite{Loureiro}. We found that the
scaling functions are fully determined by the type of correlations
present in the initial conditions and basically of two types.  All $2
\leq q\leq 4$ clean and weak disordered Potts models quenched from
equilibrium at $T_0\to \infty$ and clean $q>4$ Potts models quenched
from equilibrium at $T_c$ (the transition is of first order) share
their scaling function (see Fig.~\ref{fig:scaling-Crt}). Different
functions are obtained after quenches from $T_c$ in clean systems with
$2\leq q \leq 4$ and dirty systems with $q>4$; in all these cases the
transition is of second order and the critical exponents, in
particular $\eta$, depend on $q$ (although
weakly)~\cite{Jacobsen-Picco}.

On the other
hand, we analysed the geometric properties of areas and perimeters in
the $2d$ RBIM and we discuss these results in the next section.

\textcolor{black}
{
\section{Statistics and geometry of coarsening}}
\label{sec:statistics in coarsening}

In~\cite{Sicilia-PRL,Sicilia-PRE} we studied the distribution of
domain areas, areas enclosed by domain boundaries, and perimeters for
curvature-driven two-dimensional Ising-like coarsening, employing a
combination of exact analysis and numerical studies, for various
initial conditions. We showed that the number of hulls per unit area,
$n_h(A,t)\,dA$, with enclosed area in the interval $(A,A+dA)$, is
described, for a disordered initial condition, by the scaling function
\begin{equation}
n_h(A,t) = 2c_h/(A + \lambda_h t)^2
\; ,
\label{eq:hull-areas}
\end{equation}
where $c_h=1/8\pi\sqrt{3} \approx 0.023$ is a universal constant and
$\lambda_h$ is a material parameter. For a critical initial condition,
the same form is obtained, with the same $\lambda_h$ but with $c_h$
replaced by $c_h/2$. For the distribution of domain areas, we argued
that the corresponding scaling function have the form 
\begin{equation}
n_d(A,t) = (2)c_d (\lambda_d t)^{\tau-2}/(A +
\lambda_d t)^{\tau}
\; , 
\label{eq:nd}
\end{equation} where $c_d$ and $\lambda_d$ are numerically
very close to $c_h$ and $\lambda_h$ respectively, and the exponent
$\tau$ is the one characterising the distribution of initial
structures, critical percolation or critical Ising. These
results were extended to describe the number density of the length of
hulls and domain walls surrounding connected clusters of aligned
spins.  These predictions were supported by extensive numerical
simulations. We also studied numerically the geometric properties of
the boundaries and areas.

\begin{figure}
\begin{center}
\includegraphics[width=6cm]{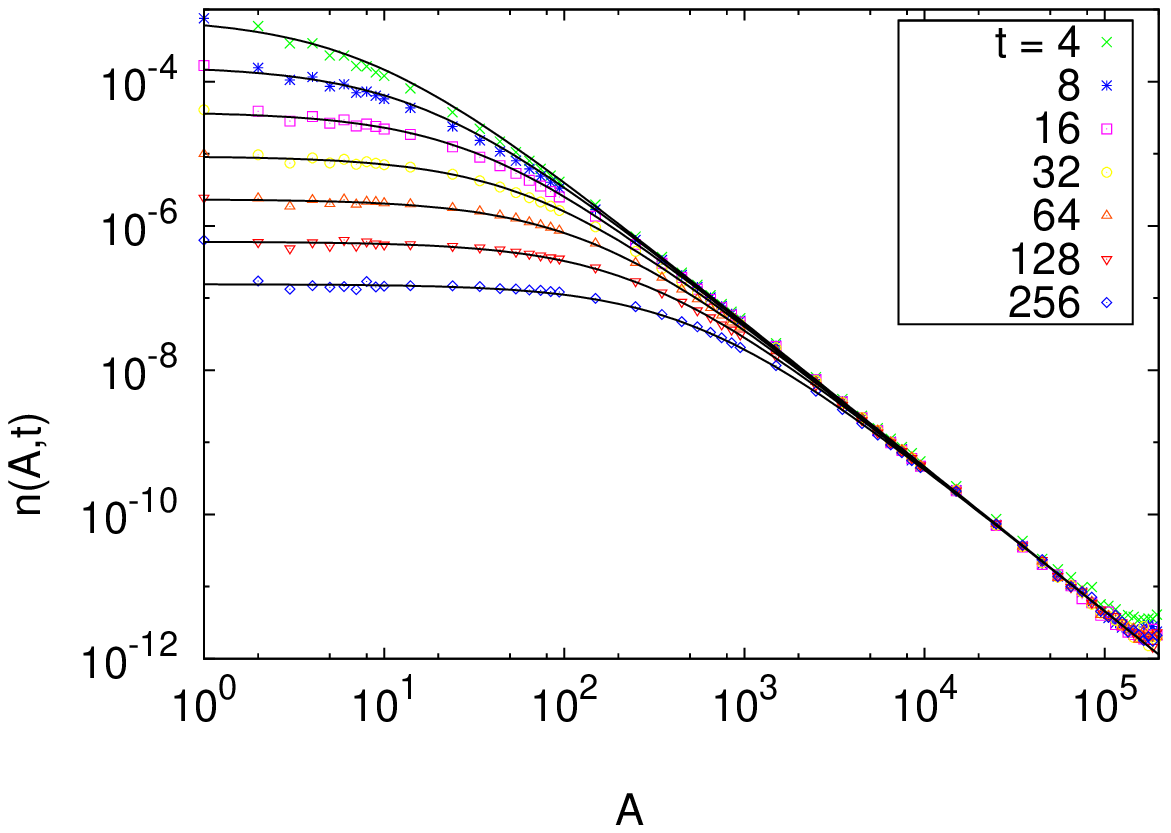}
\includegraphics[width=6cm]{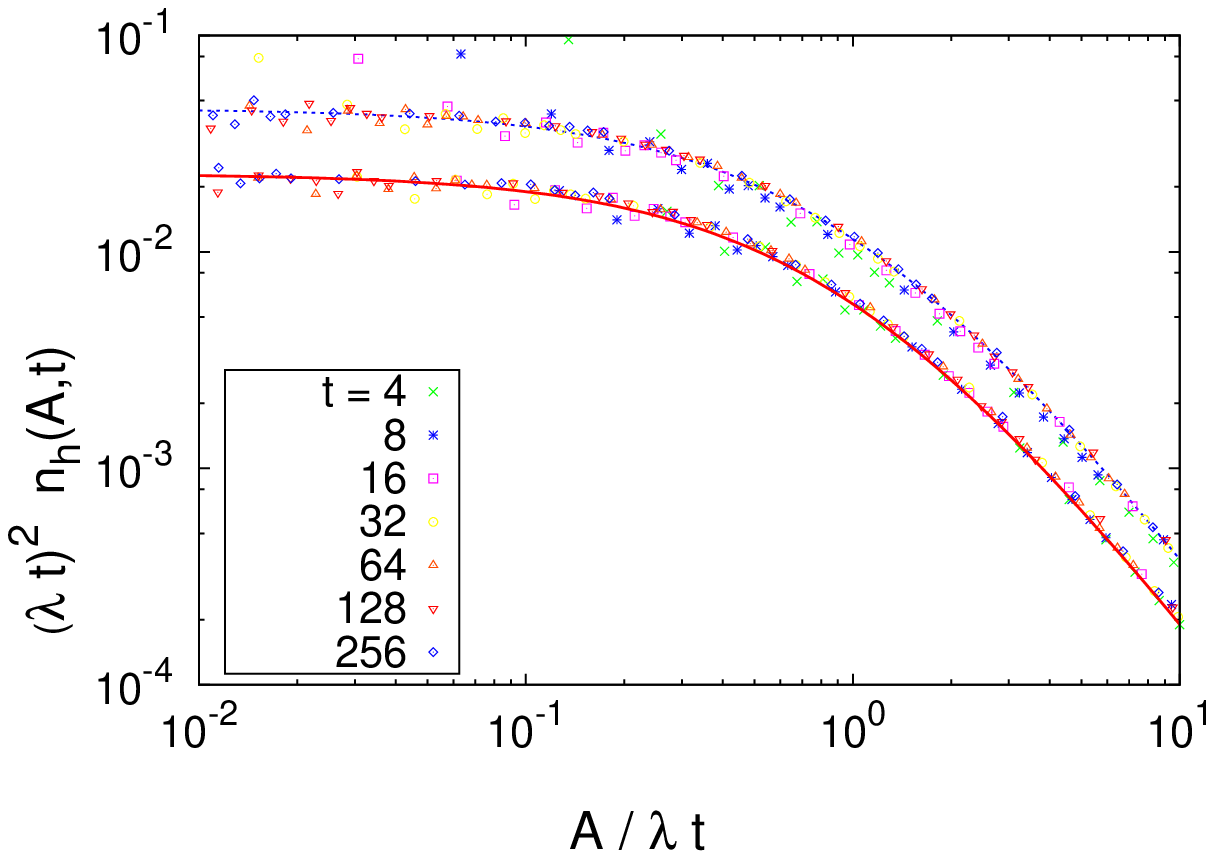}
\end{center}
\caption{Left: number density of hull-enclosed areas at different
times after a quench from $T_0\to\infty$ to $T=0$ in the $2d$IM;
numerical results are shown with points and the analytic prediction
with lines. There is only one fitting parameter, $\lambda$, that has 
been independently determined by a study of the space-time correlation. 
Right: comparison between the scaled number-density after
a quench from $T_0\to\infty$ (above) and $T_0=T_c$ (below). 
Figure taken from~\cite{Sicilia-PRE}.}
\end{figure}

The derivation of eq.~(\ref{eq:hull-areas}) is very easy. 
The number density of hull enclosed or domain areas at time $t$ as a
function of their initial distribution is
\begin{equation}
n(A,t) = \int_0^\infty dA_i \; \delta(A-A(t,A_i)) \, n(A_i,t_i) 
\;
\label{eq:integration} 
\end{equation}
with $A_i$ the initial area and $n(A_i,t_i)$ their number 
distribution at the initial time $t_i$. $A(t,A_i)$ is the
hull enclosed/domain area, at time $t$, having started from
an area $A_i$ at time $t_i$. The number 
density of hull-enclosed areas in critical percolation and 
critical Ising conditions was obtained by Cardy and Ziff~\cite{Cardy-Ziff}
\begin{eqnarray}
n_h(A,0) \sim \left\{
\begin{tabular}{ll}
 $2c_h/A^2 \;$ , & \qquad \mbox{critical percolation}, \\
 $c_h/A^2 \;$ , & \qquad \mbox{critical Ising}.
\end{tabular}
\right.
\label{eq:nhpercolation}
\end{eqnarray}
These results are valid for $A_0 \ll A \ll L^2$, with $A_0$ a
microscopic area and $L^2$ the system size. Note also that we are
taking an extra factor 2 arising from the fact that there are two
types of hull enclosed areas, corresponding to the two phases, while
the Cardy-Ziff result accounts only for clusters of occupied sites
(and not clusters of unoccupied sites).  $n_h(A,0)\,dA$ is the number
density of hulls per unit area with enclosed area in the interval
$(A,A+dA)$ (we keep the notation to be used later and set $t=0$).  The
adimensional constant $c_h$ is a universal quantity that takes a very
small value: $c_h = 1/(8\pi\sqrt{3})\approx 0.022972$.
The time dependent area is given by the Allen-Cahn result, 
$A(t) = A_i -\lambda_h (t-t_i)$. By simple integration of 
eq.~(\ref{eq:integration}) one obtains eq.~(\ref{eq:hull-areas}).

It is interesting to note that the infinite temperature initial condition
is not critical percolation on the square lattice. Still, it is very 
close to it an after coarse-graining $p=1/2$ becomes critical continuous 
percolation. The dynamics of the discrete model, thus, very quickly 
settle into critical percolation `initial' conditions and this determines the
distribution of large structures dynamically. This observation was
used by Barros {\it et al} to interpret freezing of $2d$ Ising models 
at very low temperatures~\cite{Barros}.

The calculation of the number density of domain areas cannot be done
analytically but a mean-field-like approximation that uses an expansion 
in powers of $c_h$ yields eq.~(\ref{eq:nd}), a result that is 
very convincing since it agrees well 
with numerical simulations~\cite{Sicilia-PRL,Sicilia-PRE}.

The discussion on the geometric properties of critical objects
suggests to study the fractal properties of growing structures. 
In~\cite{Sicilia-PRL,Sicilia-PRE} we pursued this line of research by 
analysing the relation between areas and perimeters during coarsening.
We summarize the results below.

Several extensions of these results are: to the Ising model with
conserved order parameter~\cite{Sarrazin}, the random bond Ising
model~\cite{Sicilia-EPL}, the Potts model~\cite{Loureiro}, and coupled map
lattices~\cite{Katzav}.

\textcolor{black}{
\subsection{General picture}
}

The summary of results in, and picture emerging from, the study of
coarsening in $2d$ systems with Ising
symmetry is
the following.

(i) We {\it proved} scaling for the number density of hull-enclosed
areas in bi-dimensional zero-temperature 
curvature-driven coarsening in the Ising universality class~\cite{Sicilia-PRL}.

(ii) We argued that in these systems temperature effects are two-fold: they 
renormalize the pre-factor in the growth law and they introduce 
thermal domains that are distributed as in equilibrium. Their 
contribution to the total distribution of structures can 
be safely subtracted~\cite{Sicilia-PRE}.

(iii) We obtained approximate expressions for the number density of
domain areas and perimeters in curvature-driven coarsening in the $2d$
Ising universality class by using a mean-field-like analytic 
argument~\cite{Sicilia-PRE}.

(iv) We checked results (i)-(iii) experimentally by 
analyzing the dynamics of a $2d$ liquid crystal~\cite{Sicilia-exp}.

(v) We derived the scaling functions of the number density of
hull-enclosed areas, domain areas and perimeters for systems in the
universality class of the $2d$ Ising model class with conserved
order-parameter dynamics quenched from high-temperature by assuming
that small structures behave roughly independently of each
other~\cite{Sarrazin}.

(vi) We analysed the area number densities in the RBIM and we found that 
super-scaling holds once the growing length is modified to incorporate
the slowing down due to pinning~\cite{Sicilia-EPL}. 

All these results are described by the following conjecture
for the number density of domains and hull-enclosed areas~\cite{Sarrazin} 
\begin{equation}
R^{2\tau}(t) \; n_{d,h}(A,t) 
= 
\frac{ (2)c_{d,h} \ a^{2(\tau-2)} 
\left[\displaystyle \frac{A}{R^2(t)} \right]^{1/2} }
{
\left\{
1+ \left[\displaystyle \frac{A}{R^2(t)} \right]^{z_d/2}
\right\}^{(2\tau+1)/z_d}
} 
\; 
\label{eq:nh-allA-super}
\end{equation}
(this expression is exact for hull-enclosed areas in curvature-driven
coarsening).  $z_d$ is the dynamic exponent (in the RBIM case, it is
the effective temperature-dependent one, $\overline z_d$). $\tau$ characterises the
initial distribution.  A way to derive
eq.~(\ref{eq:nh-allA-super}) is to assume that each time-dependent
area is independently linked to its initial value by
\begin{equation}
A^{z_d/2}(t) \approx {A_i}^{z_d/2} - R(t)
\; ,
\label{eq:super-universality} 
\end{equation}

In all these cases the expressions that we obtained have two distinct
limiting regimes.  

(vii) For areas that are much smaller than the characteristic area,
$R^2(t)$, the distributions `feel' the microscopic dynamics and
thermal agitation while for areas that are much larger than $R^2(t)$
the distributions are simply those of the initial condition. For
conserved order parameter dynamics the
Lifshitz-Slyozov-Wagner~\cite{Lifshitz} behaviour is recovered after a
quench from $T_0\to\infty$ and evolving at sufficiently low $T$.  For
critical Ising initial conditions the distribution of small areas does
not approach the Lifshitz-Slyozov-Wagner. We conjectured that the
reason is that our starting assumption, independence of domain wall
motion for small domains, is not valid due to strong correlations in
this case~\cite{Sarrazin}.

(viii) The distribution of the time-dependent 
areas that are larger than $R^2(t)$ are, in all cases, the ones of
critical continuous percolation (for all initial conditions
in equilibrium at $T_0>T_c$) and critical Ising ($T_0=T_c$).  

We also studied the geometric properties and fractal dimensions of 
different objects during coarsening. We found that

(ix) Small structures are compact and tend to have smooth boundaries,
that are pretty close to circular ($A\sim p^2$) in the conserved
order-parameter case and a little bit less so for non-conserved
order-parameter dynamics.

(x) The long interfaces retain the fractal geometry imposed by the
equilibrium initial condition.

In~\cite{Sicilia-exp} we performed 
experiments in a $2d$ liquid crystal that evolves
through the formation of domains of two chiralities. The analysis of 
data confirmed that the dynamics is of curvature-driven type, in the 
sense that the correlation functions scale with $R\sim t^{1/2}$ and, 
moreover, after a careful analysis of noise originated in the 
data acquisition procedure, we obtained a hull-enclosed area distribution 
function that agrees well with the theoretical prediction.

The geometrical analysis of coarsening cells in the $2d$ Potts model
will be presented in~\cite{Loureiro}. One might expect that other 
dynamic systems, once in the Ising non-conserved order parameter
universality class, such as some voter models~\cite{voter}, should have similar
area distributions. The case of a $2d$ Edwards-Anderson model has not been
analyzed yet.

As future work we plan to investigate how these results 
extend to $3d$ systems by analyzing volume distributions as well as 
area distributions on $2d$ cuts. Three dimensional phase separation of a 
binary mixture is realized by different physical systems, in particular
the one studied in~\cite{Vandembroucq}, that promises to be a 
nice material where to confront the results of analytics and simulations
to experiments.

\textcolor{black}
{
\section{Coarsening in the spiral model}}
\label{sec:coarsening-spiral}

Among other features, in~\cite{Corberi} we studied the dynamics of the 
spiral model after
a sudden quench, in which a completely empty configuration is evolved
from time $t=0$ onwards with the dynamic rule specified by a different
value of $p>p_0$.  This procedure is similar to the temperature quench
of a liquid.  For $p<p_c$, after a non equilibrium transient the
system attains a non-blocked equilibrium state.  The relaxation time
for attaining such a state diverges as $p\to p_c$. In the critical
case with $p=p_c$, the system approaches the blocked equilibrium state
by means of an aging dynamics similar to that observed in critical
quenches of ferromagnetic models. Freezing is not observed because the
dynamic density $\rho (t)$ is always smaller than the critical one
$\rho _c =p_c$, at any finite time.  Something different happens for
filling at $p>p_c$.  The system keeps evolving but a blocked state is
never observed despite the fact that the density of particles exceeds
$\rho _c=p_c$ at long enough times. This is no surprise since the
dynamics are fully reversible: for any $p$, each configuration reached
dynamically can always evolve back to the initial empty state by the
time reversed process, although with a very low
probability. Therefore, a blocked state cannot be dynamically
connected to the initial empty state. The
dynamics of the spiral model at $p>p_c$ strongly resembles coarsening
in ferromagnets, as illustrated in the centre and right panels in
Fig.~\ref{figsketch}. In the $p\simeq 1$ limit in which the dynamics
can be analyzed semi-analytically the system coarsens by forming
longer and longer one-dimensional objects of vacancies that are almost
frozen but not completely since they could be destroyed by the
boundaries.  The density of these objects is irrelevant in the large
size limit and the density of particles can asymptotically approach
$1$ although the system is never blocked.

\vspace{1cm}

\noindent{\it Acknowledgments} I wish to thank J. J. Arenzon, C. Aron,
M. Baity-Jesi, 
G. Biroli, A. J. Bray, S. Bustingorry, C. Chamon, F. Corberi, A. Jelic, J. L.
Iguain, A. B. Kolton, M. P. Loureiro, M. Picco, Y. Sarrazin and
A. Sicilia for our collaboration on phase ordering problems
that lead to the new results discussed in these notes. I also wish to 
thank E. Domany, P. Krapivsky, D. Vandembroucq, and F. V\'azquez for 
recent 
discussions on this problem.

\end{document}